\documentclass[prb,aps,twocolumn,10pt]{revtex4-2}
\usepackage[dvipdfmx]{graphics} 
\usepackage{epsf,graphics,graphicx,amsmath,color}
\usepackage[hidelinks]{hyperref}
\begin{document}

\title{Backbone Mediated Electrical Transport in a Double-Stranded DNA}

\author{Sourav Kundu}

\email{sourav.kunduphy@gmail.com}

\affiliation{Department of Physical Sciences, Indian Institute of Science Education and Research Kolkata, Mohanpur, West Bengal - 741246, India}

\author{Siddhartha Lal}

\email{slal@iiserkol.ac.in}

\affiliation{Department of Physical Sciences, Indian Institute of Science Education and Research Kolkata, Mohanpur, West Bengal - 741246, India}

\begin{abstract}

In the field of DNA nanotechnology, it is common wisdom that charge transport occurs through the $\pi$ stacked bases of double-stranded DNA. However, recent experimental findings by Zhuravel {\it{et. al.}} [Nat. Nanotech. \textbf{15}, 836 (2020)] suggest that electronic transport happens through the backbone channels instead of $\pi$-$\pi$ interaction of the nitrogen bases. These new experimental insights call for a detail investigation. In keeping with this, we examine charge transport properties of three characteristic double-stranded DNA sequences (periodic GC, periodic AT and random ATGC sequences) within a tight-binding framework where  backbones form the main conduction channels. Using techniques based on the Green’s function method, we inspect the single-particle density of states and localization properties of DNA in the presence of discontinuities (``nicks") along the backbone channels. We also investigate the effect of these nicks on current-voltage response using the Landauer-Büttiker formalism for a two-terminal geometry where the source electrode is attached to one backbone strand and the drain to the other. We observe that the periodic DNA sequence of GC bases is metallic in nature, while the periodic AT sequence and the random ATGC sequence are insulating. Further, the effects of nicks on the transport properties of the periodic GC sequence is interesting: while a single nick on the upper backbone does not affect electronic transport, the addition of a second nick on the lower backbone causes the current to vanish altogether. This is found to be robust against changes in the positions of the nicks, as well as the alternation of the source and drain electrodes. Analysis of the position dependence of the probability distribution of the zero-energy electronic wavefunction leads us to conclude that the insulation mechanism arises from a novel quantum interference of the electronic wavefunction from the two nicks.
\end{abstract} 


\maketitle

\section{Introduction}

Since the late 90's, biomolecules have garnered massive interests from different scientific communities because of their possible applications in nanoelectronics and spintronics~\cite{endres,zutic,genereux,cordes}. DNA, the building block of life, has traditionally been a major focal point in this regard~\cite{kelley,fink,porath,cai,tran}. DNA has three main components: nitrogen bases, pentose sugar (de-oxyribose) and phosphate groups. Of these, the pentose sugar and phosphate groups form the backbone structure of the DNA. The four nitrogen bases, adenine (A), guanine (G), cytosine (C) and thymine (T), are connected to one another via Hydrogen bonds and form a ladder-like structure famously known as a double-helix~\cite{watson}. On the other hand, pentose sugars are attached to the nitrogen bases via identical C-N bonds.

Most theoretical models of charge conduction in DNA consider charge transport through the $\pi$-$\pi$ interaction between bases~\cite{roche,guti2,Cuniberti:2002,Zhong:2003,bakhshi,ladik,gcuni,sourav3}, and completely ignore transport along the backbone sites. This includes the polaron-hopping mechanism of conduction along DNA as well~\cite{singh1, singh2}. However, some striking recent experimental results~\cite{nature_porath, porath_ref24} suggest instead that charge transport occurs through the backbones rather than the through the $\pi$ stacked bases: the introduction of ``backbone channel discontinuity'' or ``nicks'' (i.e., interruptions along the backbones due to removal of a single phosphate site from one (or both) DNA strand(s)) is observed to lead to a complete vanishing of charge transport across the DNA. In the present article, our main aim is to provide a theoretical understanding of these recent experiments. In what follows, the terms ``backbone channel discontinuity'' or ``nicks'' are used in the same spirit as they were introduced in Ref.~\cite{nature_porath}. 

We model charge transport in a double stranded DNA (ds-DNA) within a tight-binding framework, and where the backbones form the main conduction channels. While the approach taken here is related to the dangling backbone ladder model (DBLM) of DNA~\cite{sourav2}, the inclusion of the backbone conduction mechanism is novel. Theoretical investigations were performed on three different geometries (pristine, a single nick in one strand, and one nick in each of the two strands of the double-helix) for each of three DNA sequences, i.e., two periodic ds-DNA sequences (e.g., poly(dA)-poly(dT) and poly(dG)-poly(dC)) as well as a random sequence containing both A-T and G-C base pairs. First, we compute the local density of states (LDOS) profile of the three geometries in each of the three sequences. We then study transport properties for all these cases, including computations of transmission profiles and localization length with varying energy ($E$) as well as current - voltage ($I$-$V$) responses. 

Our main findings are as follows. First, we observed that the poly(dG)-poly(dC) sequence is metallic in nature, while the poly(dA)-poly(dT) and random ATGC pristine sequences are insulating. This nature prevails even in presence of a single nick in one of the strands of the double-helix. However, upon the introduction of a single nick in each of the two strands of the double-helix, the current becomes vanishingly small in all three sequences irrespective of the electrical response in their pristine condition. Further, this behaviour is observed to be independent of electrode position, as well as the position of the nicks along the two strands of the DNA double-helix. This strongly supports recent experimental observations that suggest backbone channel dominated charge transport in DNA~\cite{nature_porath, porath_ref24}. This paper is organized as follows. In Sec. II we introduce the model Hamiltonian and briefly describe our theoretical formulation. We analyze our numerical results in Sec. III, and conclude in Sec. IV.

\section{Model and Theoretical Formulation}

We employ a tight-binding framework to theoretically model charge transport along a ds-DNA. The effective Hamiltonian for the model can be expressed as follows (see schematic representation in Fig. 1)

\begin{figure*}[htb]
\centering
\begin{tabular}{c}
\includegraphics[width=0.42\textwidth]{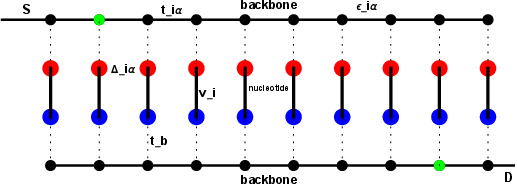}\\\\
\end{tabular}
\caption{(Color online). Schematic view of the modified ladder model representing a ds-DNA incorporating backbone conduction channels (the two solid lines on the upper and lower chains). Black dots along the two solid lines are backbone sites. Blue and Red dots are the nitrogen bases connected via Hydrogen bonds (solid black lines). Dotted lines are the identical C-N bonds between a backbone site and a nitrogen base. S and D represent source and drain electrodes respectively. The two Green dots represent introduced discontinuations (or ``nicks") in the two backbones. Various on-site energies and hopping amplitudes are shown on the respective bonds, backbone and base sites.}
\label{fig:1}
\end{figure*}
 
\begin{eqnarray}
&\hspace*{-1.0cm} H_{DNA}&= \sum\limits_{i=1,\sigma}^N\sum\limits_{\alpha=I,II}\left(\epsilon_{i\alpha}
c^{(\alpha)\dagger}_{i,\sigma}c_{i,\sigma}^{(\alpha)}
+t_{i\alpha}(c^{(\alpha)\dagger}_{i,\sigma}c_{i+1,\sigma}^{(\alpha)}+\mbox{h.c.}) \right)\nonumber \\
&&~~~+ \sum_{i=1,\sigma}^N  \left( \Delta_{i\alpha}
d^{(\alpha)\dagger}_{i,\sigma}d_{i,\sigma}^{(\alpha)}+ {v_i} (d^{(I)\dagger}_{i,\sigma}d_{i,\sigma}^{(II)}+\mbox{h.c.}) \right)\nonumber \\
&&~~~~~~+ \sum_{i=1}^N\sum\limits_{\alpha=I,II} {t_b} \left(c^{(\alpha)\dagger}_{i,\sigma}d_{i,\sigma}^{(\alpha)}+\mbox{h.c.} 
\right)~,
\end{eqnarray} 
where $c_{i,\sigma}^{(\alpha)\dagger}$ ($c_{i,\sigma}^{(\alpha)}$) and $d_{i,\sigma}^{(\alpha)\dagger}$ ($d_{i,\sigma}^{(\alpha)}$) are the fermionic creation (annihilation) operators applied at the $i$-th site of the $\alpha=I/II$ (upper/lower) backbone strands and inner  nitrogen bases respectively (see Fig.\ref{fig:1}) and $\sigma=\uparrow,\downarrow$ refers to the spin index of the electrons. $\epsilon_{i\alpha}$ and $\Delta_{i\alpha}$ represent the on-site energies for the backbones and bases respectively. $t_{i\alpha}$ represents the intra-strand hopping amplitude between two neighbouring backbones, $t_b$ the hopping amplitude between a backbone and the adjacent inner base and $v_i$ the vertical hopping between two nitrogen bases with the same site index. 

In order to explore the transport properties  of the ds-DNA, we employ source (S) and drain (D) electrodes modelled as semi-infinite one-dimensional chains that are connected to the DNA at the left and right ends respectively. The Hamiltonian of the entire system is thus given by 
\begin{equation}
H = H_{\mathrm{DNA}} + H_{\mathrm{S}} + H_{\mathrm{D}} + H_{\mathrm{tunneling}}~,
\end{equation}
where the explicit form of $H_{\mathrm{S}}$, $H_{\mathrm{D}}$ and $H_{\mathrm{tunneling}}$ are given by
\begin{eqnarray}
& H_{\mathrm{S}} & =\sum\limits_{i=-\infty,\sigma}^0\left(\epsilon c^\dagger_{i,\sigma}c_{i,\sigma}+
t c^\dagger_{i+1,\sigma}c_{i,\sigma}+\mbox{h.c.} \right)~, \\
& H_{\mathrm{D}} & =\sum\limits_{i=N+1,\sigma}^\infty\left(\epsilon c^\dagger_{i,\sigma}c_{i,\sigma}+
t c^\dagger_{i+1,\sigma}c_{i,\sigma}+\mbox{h.c.} \right)~, \\
& H_{\mathrm{tunneling}}& = \tau \sum\limits_{\sigma}\left(c^\dagger_{0,\sigma}c_{1,\sigma}^{(I)}+c^{(II)\dagger}_{N,\sigma}c_{N+1,\sigma} +
\mbox{h.c.}\right)~,
\end{eqnarray}
and where $\tau$ is the tunneling matrix element between the DNA and the electrodes.         

We use the Green's function approach to calculate the transmission probability $T(E)$ of electrons~\cite{datta1} through the DNA double-helix in this two-terminal geometry. The single-particle retarded Green's function operator representing the complete system (i.e., ds-DNA and two semi-infinite electrodes) at an energy $E$ can be written as 
\begin{equation}
G^\mathrm{r}=(E-H+i\eta)^{-1},
\end{equation}
where $\eta\rightarrow0^+$ and 
$H$ is the Hamiltonian of the entire system. Using the Fisher-Lee relation~\cite{datta1,fisher}, the two-terminal transmission probability is then defined as 
\begin{equation}
T(E)={\mathrm{Tr}} [\Gamma_\mathrm{S} G^\mathrm{r} \Gamma_\mathrm{D} G^\mathrm{a}]~, 
\label{transm}
\end{equation}
where $E$ is the incident electron energy, $G^\mathrm{a}$ is the advanced Green's function operator and Tr represents the trace over the reduced Hilbert space spanned by only the DNA system. The coupling matrices $\Gamma_\mathrm{S/D}$ are given by
\begin{eqnarray}
\Gamma_{S(\mathrm{D})} &=& i[\Sigma^\mathrm{r}_{\mathrm{S}(\mathrm{D})}
- \Sigma^\mathrm{a}_{\mathrm{S}(\mathrm{D})}]~,~\textrm{where}\\ 
\Sigma^{\mathrm{r}(\mathrm{a})}_{S(\mathrm{D})} &=& H^\dagger_{\mathrm{tunneling}} G^{\mathrm{r}(\mathrm{a})}_{S(\mathrm{D)}} H_{\mathrm{tunneling}}~,
\end{eqnarray}
and where $G^{\mathrm{r(a)}}_{\mathrm{S}(\mathrm{D})}$ and $\Sigma^{\mathrm{r(a)}}_{\mathrm{S}(\mathrm{D})}$ represent the retarded (advanced) Green's function and self-energy respectively for the source 
(drain) electrode. The self-energies are complex-valued 
\begin{equation}
\Sigma^\mathrm{r/a}_{\mathrm{S}(\mathrm{D})}=
\Delta_{\mathrm{S(D)}} \pm i \Lambda_{\mathrm{S(D)}}~, 
\end{equation}
with $\Delta_{\mathrm{S(D)}}$ corresponding to the shift of the energy levels of DNA, and 
$\Lambda_{\mathrm{S(D)}}$ to the broadening of these levels. Further, it can be shown that the coupling matrices are related to the imaginary part of the respective self-energies
\begin{equation}
\Gamma_{\mathrm{S}(\mathrm{D})}=-2~{\mathrm{Im}} (\Sigma^\mathrm{r}_{\mathrm{S}(\mathrm{D})})~.
\end{equation} 
The effective Green's functions in eq.\eqref{transm} can then be expressed within the reduced Hilbert space of the DNA system alone and in terms of the self-energies of the source and drain electrodes 
\begin{equation}
G^\mathrm{r} = 
[G^\mathrm{a}]^\dagger = 
[E-H_{\mathrm{DNA}}-
\Sigma^\mathrm{r}_S-\Sigma^\mathrm{r}_D+i\eta]^{-1}~.
\end{equation}

The two-terminal Landauer conductance at temperature $T=0$ can then be computed from the transmission at the Fermi energy ($T(E=E_{F})$)~\cite{datta1} 
\begin{equation}
g = \frac{2e^2}{h}T(E_\mathrm{F})~,
\end{equation}
and the current passing through the system for an applied bias voltage $V$ can be written as~\cite{datta1} 
\begin{equation}
I(V) = \frac{2e}{h} 
\int^{E_\mathrm{F} + eV/2}_{E_\mathrm{F}-eV/2} T(E)dE~, 
\end{equation}
where the Fermi energy has been set at $E_\mathrm{F}$=0 eV. We have assumed here that the entire voltage drop occurs only at the boundaries of the conductor. The electronic density of states (DOS) $\rho(E)$ of the DNA system can also be obtained using the Green's function formalism
\begin{equation}
\rho(E) = - \frac{1}{\pi} {\mathrm{Im[Tr}\left(G_{DNA}(E)\right)]}~,
\end{equation}
where
$G_{DNA}(E)= (E-H_{\mathrm{DNA}}+i\eta)^{-1}$ is the Green's function for the DNA system with electron energy $E$ as $\eta\rightarrow0^+$, $H_{\mathrm{DNA}}$ is the Hamiltonian of DNA, and $\mathrm{Im}$ and $\mathrm{Tr}$ represents imaginary part and trace over the entire Hilbert space respectively. 

Finally, we define the localization length ($l$) of the DNA system as the inverse of the Lyapunov exponent $\gamma$~\cite{ventra}, i.e.,
\begin{equation}
l = 1/\gamma = -\lim\limits_{L\to\infty}\frac{L}{\langle \ln(T(E)) \rangle}~,
\end{equation}
where $L$ is the length of the DNA chain in terms of backbone sites along any one chain and $\langle \quad \rangle $ denotes the average over different configurations of the nicks placed along the backbone chains. 

\section{Results and Discussions}
In keeping with the 100 hase-pair length DNA section studied in Ref.~\cite{nature_porath}, we set system length $L=100$. For our numerical calculations, the on-site energies of all nitrogen bases ($\Delta_{i\alpha}$) are taken as their ionization potentials, with the following numerical values~\cite{yan,senth}: $\Delta_\mathrm{G} = 8.177$ eV, $\Delta_\mathrm{C} = 9.722$ eV, $\Delta_\mathrm{A} = 8.631$ eV, $\Delta_\mathrm{T} = 9.464$ eV. We set on-site energies of the backbones as $\epsilon_{i\alpha}=0$~. Based on first-principle calculations~\cite{yan,senth}, we take the vertical (i.e., inter-strand) hopping to be $v_i= 0.055$ eV for GC base-pair and $v_i= 0.047$ eV for AT pair. The nearest neighbor hopping parameter in a given strand is taken as $t_{ij}= 0.35$ eV. The coupling between a nitrogen base and associated backbone ($t_b$) is taken as 0.7 eV~\cite{Cuni2002, sourav1}. In order to obtain the two geometries with nicks, we introduced them in the following fashion: the first nick is put on the upper backbone-chain at the 25th position from the source (S) end. The second nick is put on the lower backbone-chain at the 75th position from the source (S) end. We call this position configuration as 25 - 75. For single nick configurations, the nick is always put on the upper backbone strand. As mentioned earlier, all two nicks configurations have the two nicks put on opposite backbone-chains. 

\begin{figure*}[htb]
\centering
\begin{tabular}{ccc}
\includegraphics[width=0.32\textwidth]{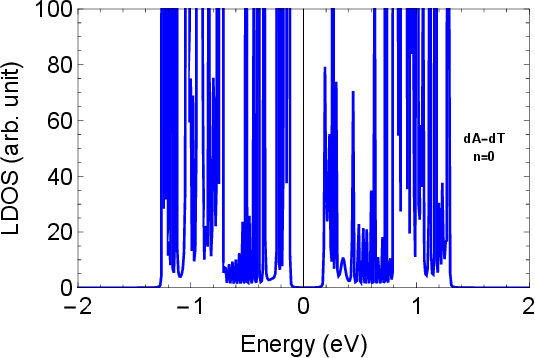}
\includegraphics[width=0.32\textwidth]{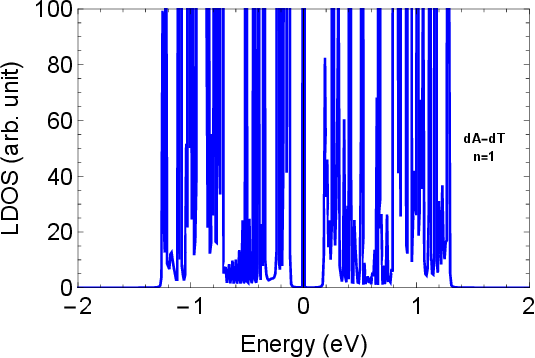}
\includegraphics[width=0.32\textwidth]{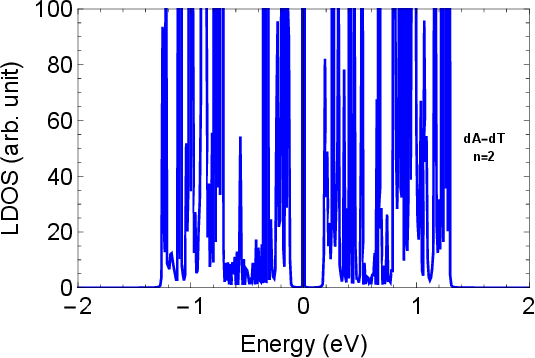}\\	
\includegraphics[width=0.32\textwidth]{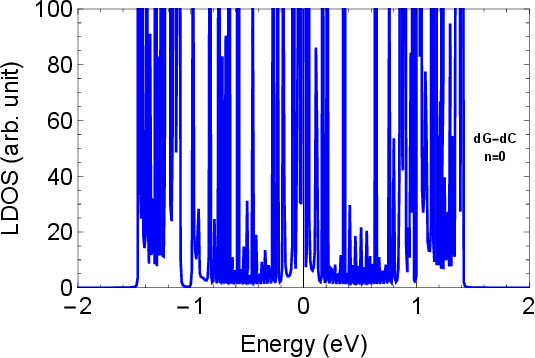}
\includegraphics[width=0.32\textwidth]{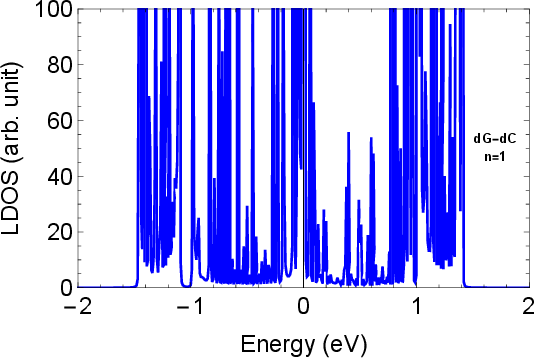}
\includegraphics[width=0.32\textwidth]{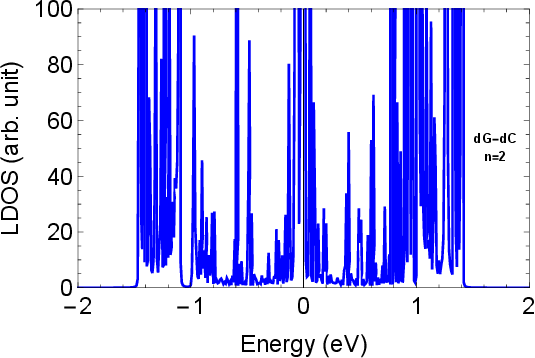}\\	
\includegraphics[width=0.32\textwidth]{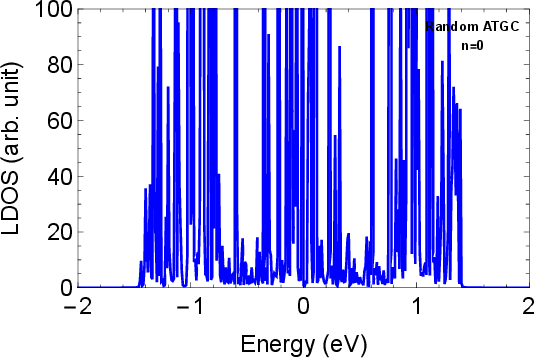}
\includegraphics[width=0.32\textwidth]{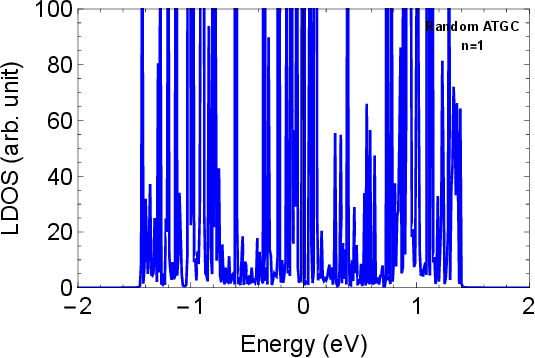}
\includegraphics[width=0.32\textwidth]{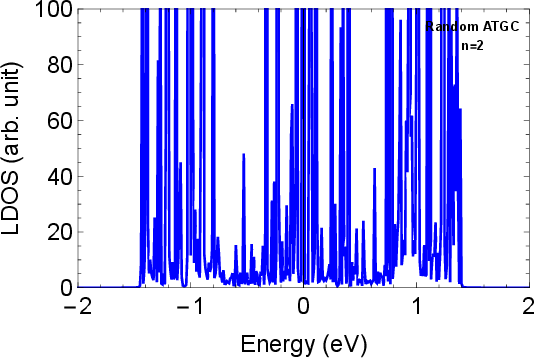}\\	
\end{tabular}
\caption{(Color online). Variation of local density of states (LDOS) with energy $E$ profiles of different sequences for three cases: pristine, with one nick and two nicks at two different strands. The three DNA sequences are poly(dA) - poly(dT) (top row):, poly(dG) - poly(dC) (center row) and Random ATGC (bottom row). The index $n$ denotes the number of nicks.}
\label{fig:2}
\end{figure*}

The single-particle density of states (DOS) profiles as a function of energy $E$ for the three DNA sequences are shown in the three rows of Fig.\ref{fig:2}. In the first row of Fig.\ref{fig:2}, we find that the pristine poly(dA) - poly(dT) sequence ($n=0$, left plot) has a gap in the DOS spectrum around $E=0$, while a narrow peak is observed in its DOS spectrum around zero (E=0) energy for both the one-nick ($n=1$, central plot) and two-nicks ($n=2$, right plot) configurations. This is indicative of the pristine poly(dA) - poly(dT) system being a band insulator. On the other hand, the reverse is observed in the poly(dG) - poly(dC) sequence (second row of Fig.\ref{fig:2}): while the pristine ($n=0$) system shows a dense spectrum of states at and around $E=0$, a reduction in DOS is observed here for the $n=1$ and, even more noticeably, the $n=2$ cases. Of the three DNA sequences, the random ATGC system (third row of Fig.\ref{fig:2}) displays the least number of changes in the single-particle DOS spectrum upon introduction of nicks. Below, we will correlate these results with our findings from investigations of transport properties.


\begin{figure*}[htb]
\centering
\begin{tabular}{ccc}
\includegraphics[width=0.32\textwidth]{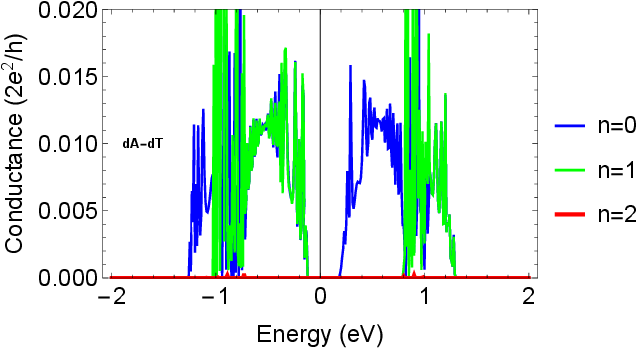}
\includegraphics[width=0.32\textwidth]{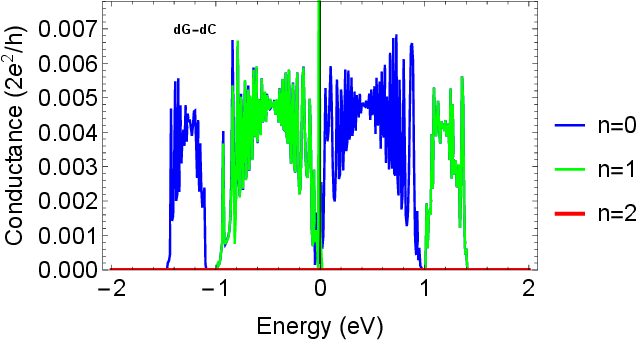}
\includegraphics[width=0.32\textwidth]{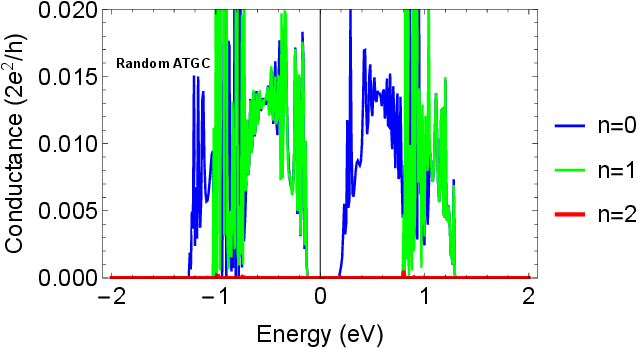}\\
\end{tabular}
\caption{(Color online). Variation of conductance $g(E)$ with energy $E$ for three DNA sequences under different conditions. See text for detailed discussion.}
\label{fig:3}
\end{figure*}

In Fig.\ref{fig:3}, we show the variation in conductance $g(E)$ with energy $E$ for the three DNA sequences. We can clearly see that while the pristine poly(dA) - poly(dT) and random ATGC sequences are insulating at vanishing bias, the pristine poly(dG) - poly(dC) sample is metallic in nature. The appearance of states in the DOS at $E=0$ upon introduction of nicks in the poly(dA) - poly(dT) does not affect the conduction, indicating that they are localised in nature. Further, the changes in the DOS of the random ATGC sequence clearly does not introduce any delocalised states. These results are expected for a band insulating poly(dA) - poly(dT) system on the other hand, and a disorder localised insulator in the random ATGC sequence on the other. We note, however, the introduction of nicks in the poly(dG) - poly(dC) system plays out differently: while a single nick on the upper backbone chain does not affect the metallic nature of the system, a second nick on the lower chain (i.e., the two nick configuration) completely suppresses all conduction. 


\begin{figure*}[htb]
\centering
\begin{tabular}{c}
\includegraphics[width=0.32\textwidth]{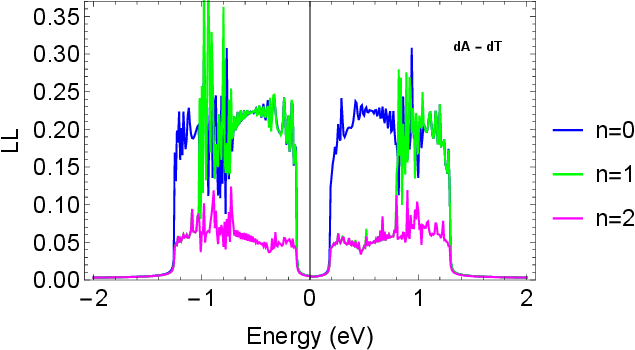}
\includegraphics[width=0.32\textwidth]{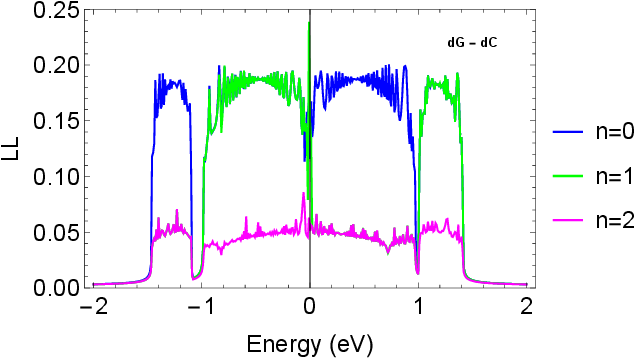}
\includegraphics[width=0.32\textwidth]{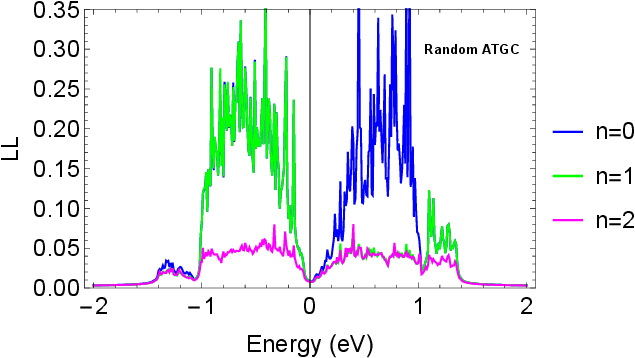}\\
\end{tabular}
\caption{(Color online). Variation of normalized localization length (LL) with energy $E$ for three DNA sequences under different conditions. See text for detailed discussion.}
\label{fig:4}
\end{figure*}

\begin{figure*}[htb]
\centering
\begin{tabular}{c}
\includegraphics[width=0.32\textwidth]{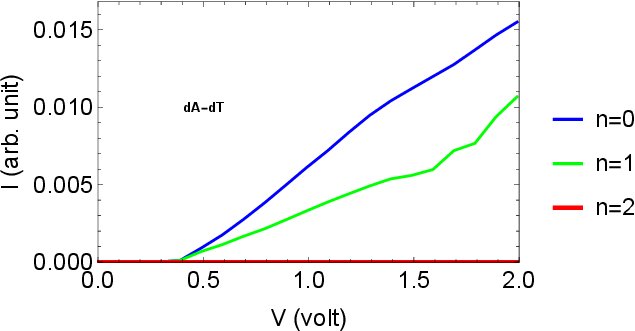}
\includegraphics[width=0.32\textwidth]{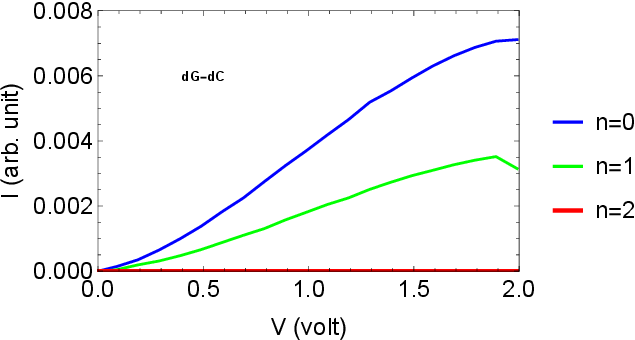}
\includegraphics[width=0.32\textwidth]{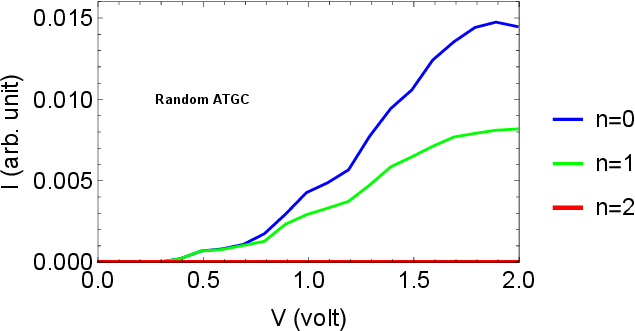}\\
\end{tabular}
\caption{(Color online). Current - Voltage (I-V) response of three DNA sequences for three configurations. See text for detailed discussion.}
\label{fig:5}
\end{figure*}

Further support of these conclusions is obtained from computations of the localisation length $l$ and the current-voltage characteristics. In Fig.\ref{fig:4}, we show the variation of the localization length ($l$) with energy $E$ for the three DNA sequences and geometries. There is no significant difference in the small $l$ at vanishing $E$ for the pristine, one nick and two nick cases for the insulating poly(dA) - poly(dT) and random ATGC sequences. On the other hand, the poly(dG) - poly(dC) sequence shows a drastic reduction in $l$ upon the introduction of the second nick, clearly indicating the strong onset of localisation in this hitherto conducting system. Similarly, the flat I-V plots at small bias shown in Fig.\ref{fig:5} for the poly(dA) - poly(dT) and random ATGC sequences clearly arise from their insulating nature and unaffected by the introduction of nicks. On the other hand, the Ohmic growth of I with V in the poly(dG) - poly(dC) sequence for the pristine and one nick cases displays the metallic nature, while the introduction of the second nick completely suppresses the current. It is interesting to note that the the I-V curves of the poly(dA) -  poly(dT) and random ATGC sequences at higher bias (i.e., beyond the insulating regime observed at small bias) also show the same qualitative response to the introduction of one and two nicks as is observed for the poly(dG) - poly(dC) sequence. This hints at the possibility that the current suppression in all three ds-DNA sequences arises from a common mechanism that affects their extended states. We have also checked our results for various two nicks configurations for all the sequences by changing the the position of the nicks (ranging from $5-95$ to $50-50$, where the first and second numbers correspond to the positions of the first and second nick from the source end placed on the upper and lower backbone strand): we find that the current becomes zero irrespective of the position of nicks. As shown in Fig.\ref{fig:7} of the Appendix, we have also checked the dependence of our conclusions for the poly(dG) - poly(dC) sequence on the positions of the source and drain electrodes: here too, we find that the current vanished identically only in case of the two-nick configuration as long as the source and drain are also attached to separate backbone strands. 


\begin{figure*}[htb]
\centering
\begin{tabular}{cc}
\includegraphics[width=0.42\textwidth]{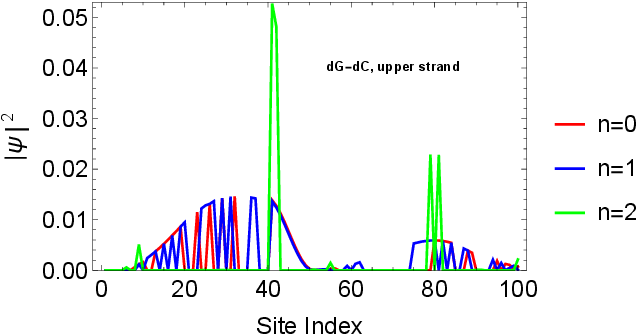}
\includegraphics[width=0.42\textwidth]{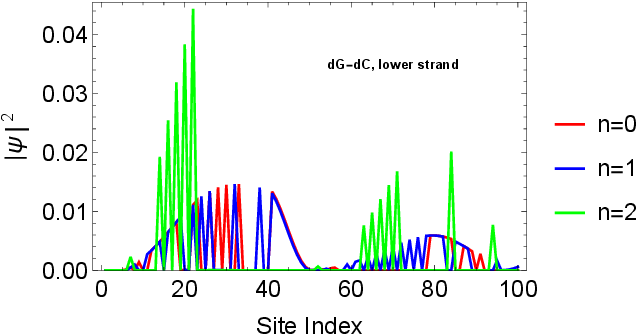}\\	
\includegraphics[width=0.42\textwidth]{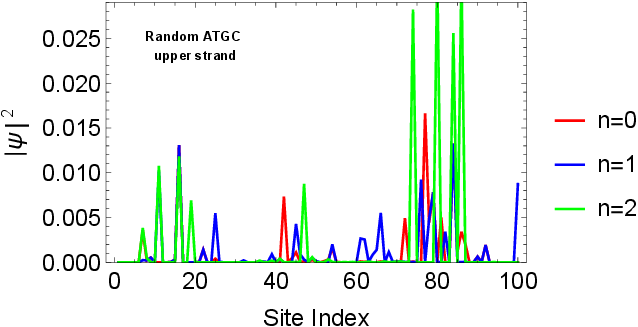}
\includegraphics[width=0.42\textwidth]{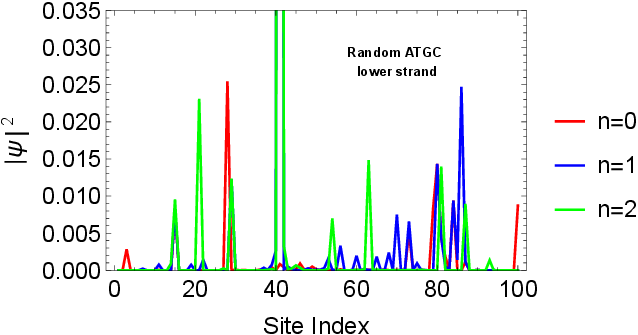}\\	
\end{tabular}
\caption{(Color online). Spatial variation of the probability density of the electronic wavefunction $|\psi|^{2}$ at $E=0$ along the two backbone chains for three DNA sequences. See text for detailed discussion.}
\label{fig:6}
\end{figure*}

Our results thus far clarify that the poly(dG) - poly(dC) sequence is quite different in its low-energy conducting properties from the other two sequences, and while it is unaffected upon the introduction of a single nick, it is affected quite drastically upon the introduction of the two nicks (one on each backbone chain). It is plausible that the low-energy extended states of the poly(dG) - poly(dC) system are only slightly affected by perturbing one of the backbones, as the other continues to facilitate conduction. However, one nick in each of the two backbones may localise these hitherto extended states through quantum interference of the electron waves between the two nicks. Should this obtain, it would drastically affect the conducting properties of the whole poly(dG) - poly(dC) sequence ds-DNA system. In order to test this hypothesis, we plot the probability distribution of the $E=0$ electron wavefunction along the sites of the two backbone chains for all cases of the poly(dG) - poly(dC) DNA sequences in Fig.\ref{fig:6} (top row); as a contrast, we have also shown results obtained from a similar analysis of all cases of the random ATGC sequence in Fig.\ref{fig:6} (bottom row).


The poly(dG) - poly(dC) sequence shows clearly the appearance of an envelope for the  probability distribution function ($|\psi|^{2}$) among the sites of both the upper and lower backbone chains (top-left and top-right panels of Fig.\ref{fig:6} respectively) for both pristine and one nick configurations. This is indicative of the fact that this state is extended along the full length of the ds-DNA. However, in presence of two nicks, the envelope is replaced by a set of peaks in $|\psi|^{2}$ that are localised at only a few points along the backbone chains. The case of the random ATGC sequence (bottom-left and bottom-right panels of Fig.\ref{fig:6}) is clearly different already: the randomness of the pristine system already renders the probability distribution function ($|\psi|^{2}$) localised among the sites of both the upper and lower backbone chains. The introduction of a single and two nicks leads only to stronger localization of the electronic wavefunction. These findings suggest strongly that the introduction of nicks at two different points on the two backbone chains leads to the localisation of the low-energy electronic wavefunction in the ds-DNA backbone chains due to quantum interference, resulting in insulating behaviour. 

\section{Concluding Remarks}
The study of charge transport through DNA has been of enduring interest, with prevalent conventional wisdom being that transport occurs due to the $\pi$ - $\pi$ bonding between the nitrogen bases of the double stranded structure. Studies on the effects of backbones and backbone-induced disorder in DNA have been conducted earlier~\cite{sourav2}, but no attention was given to the possibility of backbone-mediated transport. However, recent experiments have challenged this conventional wisdom~\cite{nature_porath}, indicating that charge propagation obtains instead via the backbone channels. Following these experimental revelations, we propose a new model to study charge transport in a ds-DNA with backbones forming the main conduction channels. Our model tight-binding Hamiltonian is based on the actual biological structure of ds-DNA, and incorporates the possibility of electrical propagation along the backbones. 

Numerical simulations of the model reveal two major findings. First among these is that a poly(dG) - poly(dC) ds-DNA sequence is metallic in nature, while a poly(dA) - poly(dT) ds-DNA sequence and a random ATGC sequence are insulating. Second, we find that a poly(dG) - poly(dC) sequence remains mostly unaffected in terms of electrical response upon the introduction of a nick in one of its backbone chains. However, introduction of two nicks (one in each of the two backbone strands) leads to a vanishing of the transmitted current across the system. This feature is observed to be quite robust against changes in the positions of the two nicks on the backbone chains, as well as the positions of the source and drain electrodes(see Appendix). We believe that the insulation mechanism arises from a novel quantum interference of the electronic wavefunction from the two nicks. Our results thus offer an explanation of recent experiments that point towards the existence of a novel transport mechanism in a ds-DNA based on conduction through the backbones. We believe that our results open up a new direction in DNA nanotechnology, and will encourage further investigations.

\section{acknowledgments}
SK thanks the CSIR, Govt. of India for funding through a Research Associate fellowship. SL thanks the SERB, Govt. of India for funding through MATRICS grant MTR/2021/000141 and Core Research Grant CRG/2021/000852.

\section{Appendix} 

\begin{figure*}[htb]
\centering
\begin{tabular}{ccc}
\includegraphics[width=0.32\textwidth]{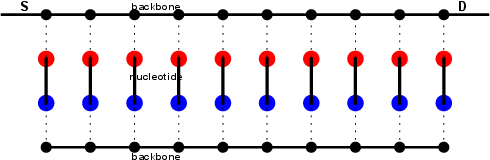}
\includegraphics[width=0.32\textwidth]{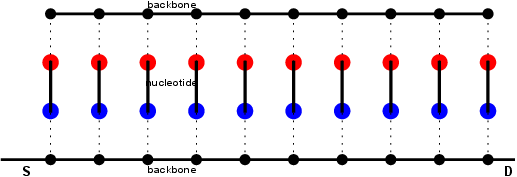}
\includegraphics[width=0.32\textwidth]{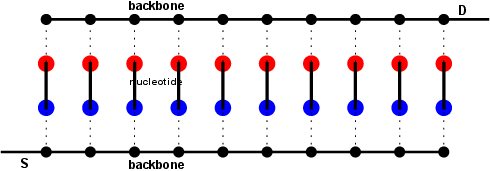}\\\\\
\\\\
\includegraphics[width=0.32\textwidth]{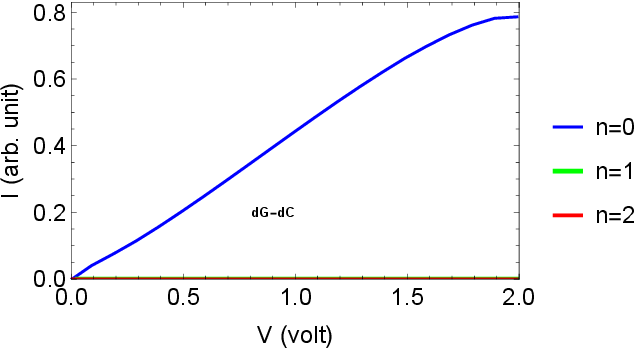}
\includegraphics[width=0.32\textwidth]{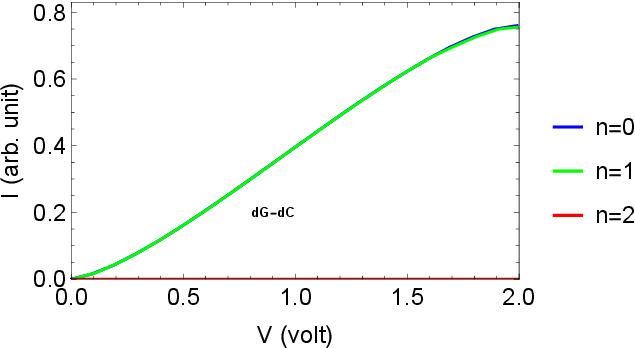}
\includegraphics[width=0.32\textwidth]{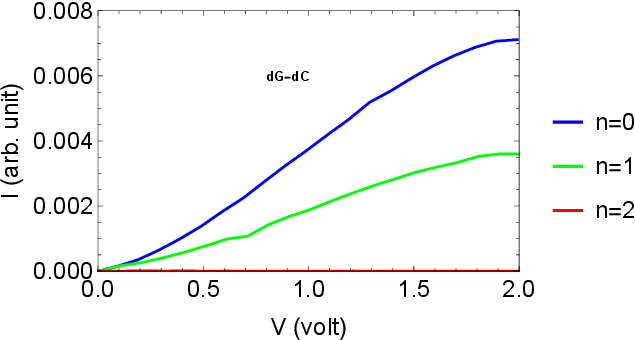}\\\
\end{tabular}
\caption{(Color online). Current - Voltage (I-V) responses of the periodic poly(dG)-poly(dC) sequence with different positions of the source (S) and drain (D) electrodes. See text for detailed discussion.}
\label{fig:7}
\end{figure*}

In Fig.\ref{fig:7}, we show the current  output for the periodic poly (dG)-poly(dC) sequence for three different electrode positions. On the left panel, both source (S) and drain (D) electrodes are attached on the upper backbone chain. Here, we find that even a single nick on the upper backbone chain leads to a vanishing current across the ds-DNA system. This likely arises from a strong localisation of all electronic wavefunction on the direct pathway between S and D.
In the middle panel, both S and D electrodes are attached to the lower backbone chain. Here, the introduction of the first nick on the upper strand leaves the transport unaffected, while that of the second nick on the lower backbone chain causes the current to vanish. Again, this is consistent with a strong localisation of all electronic wavefunction on the direct pathway between S and D. For the right panel, the S electrode is attached to the lower backbone chain and the D to the upper backbone chain (note that this is precisely opposite to the S-D configuration of Fig.\ref{fig:1}). Here, our findings of Fig.\ref{fig:5} (center panel) are repeated. This supports the possibility that the quantum interference based mechanism proposed for the insulation of the poly (dG)-poly(dC) sequence in the two nick configuration is robust as long as the S and D electrodes are on opposite backbone chains as well.

\end{document}